1

Benjamin P. Fingerhut*, Jakob Schauss, Achintya Kundu, and Thomas Elsaesser

# Aqueous Contact Ion Pairs of Phosphate Groups with Na⁺, Ca²⁺ and Mg²⁺ - Structural Discrimination by Femtosecond Infrared Spectroscopy and Molecular Dynamics Simulations

**Abstract:** The extent of contact and solvent shared ion pairs of phosphate groups with Na$^+$, Ca$^{2+}$ and Mg$^{2+}$ ions in aqueous environment and their relevance for the stability of polyanionic DNA and RNA structures is highly debated. Employing the asymmetric phosphate stretching vibration of dimethyl phosphate (DMP), a model system of the sugar-phosphate backbone of DNA and RNA, we present linear infrared, femtosecond infrared pump-probe and absorptive 2D-IR spectra that report on contact ion pair formation via the presence of blue shifted spectral signatures. Compared to the linear infrared spectra, the nonlinear spectra reveal contact ion pairs with increased sensitivity because the spectra accentuate differences in peak frequency, transition dipole moment strength, and excited state lifetime. The experimental results are corroborated by long time scale MD simulations, benchmarked by density functional simulations on phosphate-ion-water clusters. The microscopic interpretation reveals subtle structural differences of ion pairs formed by the phosphate group and the ions Na$^+$, Ca$^{2+}$ and Mg$^{2+}$. Intricate properties of the solvation shell around the phosphate group and the ion are essential to explain the experimental observations. The present work addresses a challenging to probe topic with the help of a model system and establishes new experimental data of contact ion pair formation, thereby underlining the potential of nonlinear 2D-IR spectroscopy as an analytical probe of phosphate-ion interactions in complex biological systems.

**\*Corresponding author: Benjamin P. Fingerhut,** Max-Born-Institut für Nichtlineare Optik und Kurzzeitspektroskopie, Berlin 12489, Germany, e-mail: fingerhut@mbi-berlin.de
**Jakob Schauss, Achintya Kundu, and Thomas Elsaesser:** Max-Born-Institut für Nichtlineare Optik und Kurzzeitspektroskopie, Berlin 12489, Germany.

**Running title:** Aqueous Contact Ion Pairs of Phosphate Groups

# 1 Introduction

Contact ion pairs and solvent separated ion pairs constitute fundamental solvation structures in which ions are accommodated in an aqueous environment (1,2). In particular, contact ion pairs have highest relevance, e.g., in determining the electric conductivity of liquid electrolytes and ionic aqueous solutions and in the stabilization and folding of the macromolecular biological structures of DNA and RNA (3,4) where the strong

Coulomb repulsion between the negatively charged phosphate groups of the sugar-phosphate backbone must be overcome.

The interplay of unspecific electrostatic interactions and directed solvation forces via hydrogen bonds formed with water molecules of the first solvation shell constitutes a highly complex many-body scenario and defines the relative stability of contact ion pairs which is challenging to quantify experimentally. For example, dielectric spectroscopy primarily monitors changes in the bulk water reorientation dynamics with higher sensitivity towards solvent separated ion pairs (5). Similarly, small angle X-ray scattering provides sensitivity for the globular folding state but misses microscopic details of the ion accommodating structures (6). X-ray structure determination can provide atomistic details but has a bias for immobilized ions and the water content of crystallographic samples may vary substantially from native aqueous conditions (7,8). On a qualitative level, the interaction of phosphate groups with alkali and alkaline earth ions in water has been addressed with stationary infrared and Raman spectroscopy of phosphate stretching vibrations where distinct shifts of the oscillator frequency indicate the presence of ions in vicinity of the phosphate group (9,10).

We have recently established nonlinear femtosecond infrared spectroscopy of phosphate groups as a sensitive reporter of the dynamics at the phosphate-water interface, exploiting the particularly high sensitivity of the phosphate stretching vibration towards their local hydration environment (11-16). Dimethyl phosphate (DMP) was established as valuable model system of the phosphate backbone vibrations in DNA and RNA (14,17) providing insight into solvent field fluctuation time scales and amplitudes. Femtosecond two-dimensional infrared (2D-IR) spectroscopy in conjunction with density functional simulations have allowed for a dynamic characterization of contact ion pairs of DMP with magnesium ($Mg^{2+}$), calcium ($Ca^{2+}$) and sodium ions ($Na^+$) (15,16).

Molecular dynamics (MD) simulations complement experiments and, in principle, provide an atomistic picture of ion solvation and contact ion pair formation with high computational efficiency (18,19). Nevertheless, the simulation of ion pairing via classical MD simulations relying on fixed-charge force fields is particular challenging due to uncertainties in quality of the employed force fields (20) and the required long equilibration and simulation times (21). In particular, for the highly charged divalent cations it is up to now unclear which level of theory is required for reliable description of contact ion pair structures and energetics and if the neglect of charge transfer and polarization are fundamental limitations (22,23).

Structural differences among contact ion pairs in aqueous solutions have hardly been addressed and typically been considered on the level of the charged constituents only. In this study, we present linear infrared, femtosecond infrared pump-probe and absorptive 2D-IR spectra of the asymmetric stretching vibration $v_{AS}(PO_2)^-$ of DMP in presence of the ions $Na^+$, $Ca^{2+}$ and $Mg^{2+}$, complemented by long-time MD simulations. Ion pair formation is monitored via a blue shifted feature in the vibrational spectra for which we identify characteristic differences for the different ions, reflecting the different ion charge states ($Na^+$ vs $Ca^{2+}$ and $Mg^{2+}$) and differences in ion radius for a given charge state ($Ca^{2+}$ vs $Mg^{2+}$). In particular, differences in vibrational lifetime of contact ion pairs with $Ca^{2+}$ and $Mg^{2+}$ compared to isolated DMP increase the contrast towards the contact ion pairs in the time resolved measurements making 2D-IR an excellent analytical tool for contact ion pair sensing. The observed spectral differences are traced back to structural differences of solvation shell and contact ion pair geometry as revealed by MD simulations that closely reflect findings from benchmark DMP-ion-water cluster geometries. Our results demonstrate that the intricate properties of the solvation shell around the phosphate group and the ion are essential to explain the experimental observations.

# 2 Methods

## 2.1 Experimental Methods

### 2.1.1 Sample preparation

Aqueous stock solutions were prepared by dissolving 0.2 M dimethyl phosphate sodium salt ($Na^+DMP^-$, Alfa chemistry, 97% purity) in $H_2O$ (ultra-quality, Roth), followed by adding varying amounts of sodium chloride (NaCl, Fluka), calcium chloride ($CaCl_2$ anhydrous, VWR), or magnesium chloride hexahydrate ($Mg(H_2O)_6Cl_2$, Merck) with concentrations ranging from 0.5 to 2 M. Reference samples were prepared for the ionic solutions with the same ion concentration but without DMP. In the investigated concentration range, the ionic species are fully dissociated and solvated separately (24,25).

### 2.1.2 Infrared linear absorption measurements

Linear infrared absorption spectra were measured with a commercial FTIR spectrometer (Bruker Vertex 80v). Spectra were collected in transmission mode for 0.2 M Na$^+$DMP$^-$ in H$_2$O and with the samples containing NaCl, CaCl$_2$, and Mg(H$_2$O)$_6$Cl$_2$ varying in a concentration range from 0 to 2 M. The samples were held in a commercial liquid cell (Harrick) in between two 1 mm thick BaF$_2$ windows separated by a 25 μm Teflon spacer. Infrared spectra of samples with Mg$^{2+}$ ions were measured using a 56 μm spacer and the absolute absorbance was rescaled to a 25 μm sample thickness. Reference infrared spectra of ionic solutions without addition of DMP were taken under identical conditions for each salt concentration and subtracted from the corresponding spectrum with DMP to remove the infrared background absorption. Fig. 1 (a-d) shows the respective corrected linear infrared absorption spectra of DMP for ion concentrations of 2M (black solid lines).

### 2.1.3 Pump-probe experiments

Femtosecond infrared pump-probe spectra were measured with a two-color experimental setup described in detail in Ref. (26). The output of a Ti:sapphire regenerative amplifier system (Coherent Libra, 1 kHz, 3.5 mJ, 800 nm) was split to drive two independent home-built optical parametric amplifiers (OPA) in combination with difference frequency generation (0.75 mm thick GaSe crystal) to generate independently tunable mid-infrared pulses. In the present experiments, both pulses were centered at 1220 cm$^{-1}$ with a spectral width (FWHM) of 160 cm$^{-1}$, pulse energies of 1-2 μJ and a duration of 110 fs. The pump beam is chopped at half the repetition rate (500 Hz), and focused into the sample (focal diameter ~100-150 μm). Probe pulses are attenuated by a factor of 100, passed over a delay stage, and split into two identical copies. One of these pulses is focused into the excitation volume to probe the pump induced absorbance changes. The other pulse is focused into the sample outside the excitation volume and acts as a reference to correct for shot-by-shot intensity fluctuations. After passing through the sample the probe and reference beam were detected by a dual 64-pixel mercury cadmium telluride (MCT) detector array (spectral resolution 2 cm$^{-1}$). From the measured frequency- and delay-dependent intensities I and I$_0$ for the pumped and unpumped sample respectively, the absorbance change is derived as:

$$\Delta A(T, \nu_{pr}) = -\log\left[\frac{I^{pr}(T, \nu_{pr})}{I_0^{pr}(\nu_{pr})} \cdot \frac{I_0^{ref}(\nu_{pr})}{I^{ref}(T, \nu_{pr})}\right]$$

The ultrafast pump-probe measurements were performed with samples in a home-built nanofluidic cell with two 500 nm thick silicon nitride (Si$_3$N$_4$) windows separated by a 25 μm Teflon spacer. The Si$_3$N$_4$ membranes show a negligible contribution to the nonlinear signal in the investigated 1150-1300 cm$^{-1}$ frequency range.

### 2.1.4 2D-IR experiments

Details of the 2D-IR set-up have been presented elsewhere (27). Femtosecond pulses tunable in the near-infrared are generated by three optical parametric amplification steps in beta barium borate crystals driven by a commercial Ti:sapphire laser system (repetition rate 1 kHz). Difference frequency mixing of the near-infrared signal and idler pulses in a 0.5-mm thick GaSe crystal provides mid-infrared pulses of a 115 fs duration at a center wavelength of 1215 cm$^{-1}$ (spectral width (FWHM) 150 cm$^{-1}$, energy 8 μJ). In a box-CARS beam geometry, three pulses (each ~2 μJ) focused into the sample generate a photon echo signal. This signal is heterodyned by a fourth pulse (local oscillator) traveling through the sample, dispersed by a monochromator, and detected by a 64-pixel MCT detector array (spectral resolution 2 cm$^{-1}$), thereby defining the detection frequency axis $\nu_3$. A Fourier transform of the signals along the coherence time $\tau$, the delay between the first two pulses, generates the excitation frequency $\nu_1$. The population or waiting time T is the delay between the second and third pulse. In all figures, the absorptive 2D signal, i.e., the real part of the sum of the rephasing and nonrephasing signal is plotted with a relative signal change between neighboring contour lines of 6.5 %. All 2D-IR measurements were performed using a commercial liquid cell (Harrick) in between two 1 mm thick BaF$_2$ windows with a 12 μm Teflon spacer.

## 2.2 Molecular Dynamics (MD) simulations

MD simulations were performed with the AMBER 18 software (28) employing the *ff99bsc0* force field. The initial model structure of dimethyl phosphate anion (DMP) in gg conformation was taken from Ref. (14). The starting model was placed in a truncated octahedral solvation box with a 20.0 Å buffer region. Partial charges and force field parameters of the respective DNA sugar-phosphate backbone atom types were employed for DMP. For charge neutrality, either a single Na$^+$, or Ca$^{2+}$, Mg$^{2+}$ with a single Cl$^-$ ion were added by replacing random water molecules more than 4.0 Å away from DMP. Employed water models were TIP4P-FB (29) and SPC/E.  Tested ion parameters for respective water models are Joung-Cheatham (30) ion parameters, Li-Merz 12-6 ion parameters (HFE and IOD sets) (31) and 12-6-4 Li-Merz ion parameters (32,33).

MD simulations were performed in the NPT ensemble (pressure 1.0 bar, 2 ps pressure relaxation time, Langevin dynamics with 1 ps collision frequency for temperature regulation) with a time step of 2 fs and SHAKE bond length constraints on bonds involving hydrogen atoms. Periodic boundary conditions were imposed with electrostatic interactions evaluated by the particle mesh Ewald method, employing a cut-off for long range interactions of 10.0 Å. Coordinates were written to file every 250 time steps (0.5 ps).

Equilibration was performed by initial minimization of solvent and ion molecules, restraining the atomic positions of DMP (harmonic constraints 500.0 kcal/mol Å$^{-2}$, 1000 optimization steps), followed by minimization of the entire system (2500 optimization steps). Subsequent short MD was performed to gradually heat the system to 300 K with weak harmonic position restraints on DMP (10 kcal/mol Å$^{-2}$, NVT ensemble, 20 ps simulation time). Temperature equilibration was followed by pressure equilibration in the NPT ensemble without applying position restraints for 10.0 ns. The total simulation time amounts to 11.68 μs, production runs of the two-dimensional potential of mean force (2D-PMF) cover in total 4.77 μs of simulation time (1.59 μs for Na$^+$, Ca$^{2+}$ and Mg$^{2+}$ ions, respectively). Simulations were performed with the GPU accelerated PMEMD.CUDA program (34,35) on Tesla K80 and Tesla V100 GPUs. Radial distribution functions (rdf) were calculated with the *cpptraj* program with respect to the cation, and phosphorous or oxygen atoms, of DMP. The symmetry of ion coordination to O1P and O2P atoms served as an indicator of convergence of the rdf.

# 3  Results and Discussion

## 3.1 Vibrational Lifetimes

Femtosecond pump-probe spectra of the asymmetric phosphate stretching mode $v_{AS}(PO_2)^-$ of DMP (c=0.2 M) and DMP with a 2 M ion excess concentration are presented in Fig. 1 (a-d) for different pump-probe delays in order to determine vibrational lifetimes of the $v_{AS}(PO_2)^-$ mode subject to presence of different ions. The pump-probe spectra are shown together with the respective linear absorption spectra (black solid lines), the spectrum of pump pulse is indicated in Fig. 1(a) as the dashed black line.  The linear absorption spectra of DMP with Na$^+$ ions are almost indistinguishable from the linear absorption spectra of DMP (Fig. 1a and b). The lineshape of the linear $v_{AS}(PO_2)^-$ absorption band of the DMP reference sample exhibits a plateau-like part around the maximum which arises from contributions of the gg and gt DMP conformers (14). The linear absorption spectrum of DMP with Ca$^{2+}$ ions (Fig. 1c) is broadened and shows a noticeable blue-shifted shoulder around ~ 1240 cm$^{-1}$ arising from contact ion pair formation (16). In the presence of Mg$^{2+}$ ions (Fig. 1d) this feature is moderately amplified and shifted to ~ 1250 cm$^{-1}$.

The prominent negative signal in femtosecond pump-probe spectra centered at 1220 cm$^{-1}$ is due to bleaching and stimulated emission signal on the v=0→1 transition and the positive signal centered at 1170 cm$^{-1}$ is due to the anharmonically red-shifted v=1→2 transition of the $v_{AS}(PO_2)^-$ vibration. The pump-probe spectra of the DMP reference sample and DMP with Na$^+$ ions are almost indistinguishable in signal shape, signal amplitude and temporal evolution of the signal amplitude (Fig. 1a,b). The respective pump-probe spectra in presence of Ca$^{2+}$ or Mg$^{2+}$ ions are substantially broadened towards higher frequencies. The re-duced amplitude of the bleaching and stimulated emission signal compared to the DMP reference sample is rationalized via the spectral overlap of bleaching and stimulated emission signal contributions and the excited state absorption contribution to the signal arising from DMP with and without $Ca^{2+}$ ions. The amplitude of the bleaching and stimulated emission signal of DMP with $Mg^{2+}$ ions is further reduced at ~ 1220 $cm^{-1}$ compared to the DMP reference sample and a distinct contribution centered around ~ 1250 $cm^{-1}$ arising from DMP with $Mg^{2+}$ in contact is identified.

Kinetic traces for fixed probe frequencies are summarized in Fig. 2. We focus on three spectral positions, a probe frequency of 1175 $cm^{-1}$ in the range of the v=1→2 transition, a frequency of 1220 $cm^{-1}$ at the maximum negative pump-probe signal (Fig. 1a-d), and at 1250 $cm^{-1}$ in the range of the blue-shifted fundamental transitions of the contact ion pairs (Fig. 1c-d). The time evolution at positive delay times is dominated by the decay of the v=1 state of $v_{AS}(PO_2)^-$ for both DMP solvated by water and DMP contact pairs with the respective ions. While the decay at 1220 $cm^{-1}$, i.e., the frequency position of DMP solvated by water is unaffected in the different measurements, the decay at 1250 $cm^{-1}$ is slightly slower at the frequency position assigned to the DMP contact pairs. The spectrally resolved pump-probe transients in Fig. 2 (a-d) display a finite rise time at negative delay times that is caused by the perturbed free induction decay of the oscillator (13,36). The residual negative signal at large delay times (> 2 ps) is assigned to the formation of a hot ground state due to the ultrafast decay of the v=1 states of the $v_{AS}(PO_2)^-$ vibration.

The pump-probe transients were fitted with single-exponential functions (Fig. 2a-d, solid lines). The extracted population decay times are summarized in Table 1. The time constants at 1175 $cm^{-1}$ and 1220 $cm^{-1}$ are (within experimental accuracy) identical for DMP and DMP with 2 M excess ion concentration, revealing a prominent ~ 350 fs timescale of the decay of the v=1 state of the $v_{AS}(PO_2)^-$ vibration. With $Ca^{2+}$ and $Mg^{2+}$ ions present, the signal at 1250 $cm^{-1}$ in the range of the contact ion pair absorption shows a slower decay ($\tau$ = 430 – 580 fs), i.e., a longer vibrational lifetime than water-solvated DMP.

## 3.2 2D-IR Spectra

Figure 3 presents absorptive 2D-IR spectra of 0.2 M DMP in (a) $H_2O$, (b) 0.2 M DMP in $H_2O$ with 2 M $Na^+$, (c) $Ca^{2+}$, and (d) $Mg^{2+}$ ion excess concentration added to the sample solution, recorded at a waiting time T=500 fs. Absorptive 2D signals are plotted as a function of excitation frequency $v_1$ and detection frequency $v_3$. Contributions arising from ground state bleaching and stimulated emission of the v = 0 →1 transition are shown with yellow-red contours and v = 1 → 2 excited state absorption contributions are shown in blue. Signal amplitudes are normalized to the maximum ESA signal for the individual 2D spectra.

The 2D spectrum of DMP in neat $H_2O$ (Fig. 3a) displays a single peak on the v=0→1 transition. Moderate inhomogeneous broadening of the peak is evident from the elliptic lineshape that appears tilted with respect to the excitation frequency axis and is similar to data taken at T=300 fs (15). Upon addition of $Na^+$ excess ions, the v=0→1 feature of the 2D spectrum (Fig. 3b) broadens along the diagonal, as quantified from diagonal cuts presented in Fig. 3(f). For $Ca^{2+}$ and $Mg^{2+}$ excess ions (Figs. 3c,d) the broadening along the diagonal is more pronounced and an isolated blue-shifted component can be identified in the 2D spectrum, corroborating blue-shifted components that only appear as shoulders in the linear spectra (Fig. 1). Importantly, 2D cross peaks between the blue-shifted contributions to the 2D spectra and the original DMP band are absent, which shows that the underlying vibrations are uncoupled and arise from distinct chemical species. The absence of cross peaks moreover points to a negligible chemical exchange of the species underlying the different components of the 2D spectrum, i.e., the fraction of contact ions is preserved on the time scale of the experiment. Due to the longer lifetime of the asymmetric phosphate vibration $v_{AS}(PO_2)^-$ of the contact ion pair (cf. Fig. 2), a relative enhancement of the blue-shifted component of the 2D signal is observed with increasing waiting time T. The enhanced spectral separation due to lifetime differences of contact ion pair and DMP in neat $H_2O$ is particularly evident from the comparison of the cuts along the frequency diagonal (Figs. 3g,h).

## 3.3 Contact Ion Pair Geometric Structures

Microscopic insight in the different DMP-ion complexes was obtained with the help of MD simulations. Figure 4 presents radial distribution functions g(r) for a variety of tested ion and water models. The simulations of g(r) for $Na^+$ ions show a reduced tendency of $Na^+$ contact ion pair formation for the 12-6-4 Li-Merz ion parameters compared to the Young-Cheatham (Y/C) ion parameters. The 12-6-4 Li-Merz ion parameter set, where an additional attractive term is introduced to mimic charge-induced dipole interaction, shows a comparable tendency of ion pair formation as Li/Merz (HFE and IOD) ion parameters. For $Ca^{2+}$ ions we find a reduced tendency of contact ion pair formation for the 12-6-4 Li-Merz ion parameter set compared to Li/Merz (IOD) ion parameters. Comparison of TIP4P-FB and SPC/E model water models yields a slightly reduced tendency of contact ion pair formation for TIP4P-FB for 12-6-4 Li-Merz ion parameters (Fig. 4b,c). A recent study (37) on $Mg^{2+}$ ion parameters demonstrated similar good performance of the 12-6-4 Li-Merz ion parameter set.

During the ~1 µs simulation time employed for the calculation of g(r), we observe frequent and rare (~10) transformations of contact ion pairs into solvent separated ion pairs for $Na^+$ and $Ca^{2+}$, respectively. The limited number of transitions from contact to solvent-separated ion pairs for $Ca^{2+}$ poses statistical challenges for the accuracy of g(r). We thus have analyzed the radial distribution functions for different (1.0 µs) segments of a long-time MD trajectories 1.59 µs trajectory (12-6-4 Li-Merz ion parameters, TIP4P-FB water model, see below) and find that the relative population of contact and solvent-separated ion pairs agrees within 9 %. The statistical significance of g(r) was further analyzed via the symmetry of O1P…$Ca^{2+}$ and O2P…$Ca^{2+}$ contact ion pair formation. Respective differences are found to be on the order of 5 % and closely mirror the uncertainty from the different trajectory segments. Both procedures indicate an error on the order of 10 % of relative population of species.

In long-time MD trajectories (1.59 µs for $Na^+$, $Ca^{2+}$ and $Mg^{2+}$ ions), the 12-6-4 Li-Merz ion parameter set together with the TIP4P-FB water model was employed because of the moderately reduced tendency of contact ion pair formation. Figures 5(a-c) present results of long-time MD trajectories that analyze contact ion pair geometries of DMP with $Na^+$, $Ca^{2+}$ and $Mg^{2+}$ in a water surrounding. The simulations start from contact ion pair geometries generated by constraining the P…$Ion^{x+}$ distance during the first 1 ns of temperature equilibration. Contact ion pair geometries were analyzed via the two-dimensional potential of mean force (2D-PMF) for the different ions with the P…ion distance and the P…O1…ion angle $\alpha = \sphericalangle(Ion^{x+}…O1…P)$ as the relevant coordinates. (cf. Fig. 5e,f). The angle $\alpha$ takes a value of 180° for linear arrangements of the P=O group and the ion while $\alpha \approx 90$ ° if the ion is placed in the center of the bisector of the $(PO_2)^-$ group. During the trajectories, we observe frequent, rare and no transformation of contact ion pairs into solvent separated ion pairs for $Na^+$, $Ca^{2+}$ and $Mg^{2+}$ ions, respectively.

For $Na^+$ (Fig. 5a), ion pairs separated by a single $H_2O$ molecule were found to be the most stable species, being characterized by P…$Na^+$ distances of $\approx 5$ Å and $\alpha \approx 90°$. Contact ion pairs are characterized by a P…$Na^+$ distance of 3.0-3.6 Å and are slightly less stable than the solvent separated ion pairs (~ 200 cm$^{-1}$). The contact pairs show a broad distribution in the angular coordinate $\alpha$ with a center value around 133°. Contact ion pair structures characterized by a $Na^+$ ion located in the $(PO_2)^-$ bisector are slightly less stable (P…$Na^+$ distance $\approx 4$ Å and $\alpha \approx 75-90°$). Within the simulation time of 1.59 µs, the $Na^+$ ion explores the full configuration space of the 2D-PMF and the barrier of contact ion pair interconversion while transitions to solvent separated ion pairs are facilitated by thermal motion (~620 cm$^{-1}$ for the 12-6-4 Li-Merz/ TIP4P-FB model parameter combination).

For $Ca^{2+}$ (Fig. 4b), the energetic ordering of solvent separated ion pairs and contact ion pairs is reversed compared to $Na^+$ with the contact geometry being more stable by about 286 cm$^{-1}$. Due to the larger ion size the P…$Ca^{2+}$ distance increases to 3.78 Å. In the DMP…$Ca^{2+}$ contact ion pair a similar broad distribution is sampled in the angular coordinate $\alpha$ but the minimum of the distribution is shifted towards distinctly larger values ($\alpha \approx 164°$). Transition between both species and $Ca^{2+}$ relocations between solvent separated and contact ion pairs are still facilitated at moderate number (n≤10) by thermal motion within the 1.59 µs simulation time. The energy barrier in the contact ion pair to solvent separated ion pair transition, estimated via the lowest barrier transition, is ~1800 cm$^{-1}$ from the 2D-PMF with the 12-6-4 Li-Merz/ TIP4P-FB model parameter combination.

For $Mg^{2+}$ (Fig. 5c), only contact ion pairs are sampled during the 1.59 µs simulation time, in agreement with the substantial barrier separating solvent separated from contact ion pairs (38) and the microsecond water exchange times around $Mg^{2+}$ reported from NMR studies (39). Compared to $Ca^{2+}$, the P…$Mg^{2+}$ distance is substantially contracted (3.48 Å) and the angular distribution is centered around $\alpha = 167°$, a similar value as observed for $Ca^{2+}$.

In Figure 5(d), minimum energy angular profiles along $\alpha$ are compared for the DMP complexes with $Na^+$, $Ca^{2+}$ and $Mg^{2+}$. The angular energy profiles were derived from the 2D-PMF by locating the energy minima along the contact ion pair P…ion distance (< 4 Å) for each $\alpha$. For doubly charged $Ca^{2+}$ and $Mg^{2+}$ we find that the minima of the angular energy profile are located at $\alpha_{min} > 163°$ indicating an approximate linear arrangement of the P=O group and the ion. For $\alpha > \alpha_{min}$ the angular profiles of both ions appear similar while the wider shape for $Ca^{2+}$ at $\alpha < \alpha_{min}$ reflects the more fluxional and less rigid solvent shell around $Ca^{2+}$. Notably the shape of the angular profile of $Ca^{2+}$ resembles the shape of the singly charged $Na^+$ for $\alpha < 135°$ but this region of the PMF is destabilized compared to $Na^+$.

Table 2 provides a comparison of key contact ion pair geometric parameters obtained from MD simulations and DFT simulations on $DMP(H_2O)_N M^{x+}$ cluster geometries for the different ions $Na^+$, $Ca^{2+}$ and $Mg^{2+}$, DFT optimized cluster geometries are taken from (16). For P…$Ion^{x+}$ distances we find reasonable agreement (< 0.14 Å) for all investigated ions and for $\alpha$ the respective ion pair geometries agree within 11°. We note that deviations in the P…$Ion^{x+}$ distance are particularly sensitive to moderate deviations in angle $\alpha$ because direct coordination of ions occurs with the phosphate oxygen atoms. Besides the good agreement in contact ion pair geometric parameter, the MD simulations reproduce the energetics of prototypical solvation geometries found in DFT simulations on $DMP(H_2O)_N M^{x+}$ clusters (characterized by the sign of $\Delta E$ in Table 2). For the $Na^+$ ion, the most stable contact ion pair solvation geometry is characterized by the intercalation of the ion into the tetrahedral hydrogen bond geometry around the O1 atom of the $(PO_2)^-$ group (characterized by $\alpha \sim 127$-$133°$, cf. Table 2 and Fig. 5d,e). Such structures are found to be the most stable structures in both MD simulations and DFT simulations on $DMP(H_2O)_N Na^+$ clusters. For $Ca^{2+}$ and $Mg^{2+}$ ions, water molecules in the first solvation shell of the ion are replaced by an oxygen atom of the $(PO_2)^-$ group, substituting one of the water oxygen atoms in the first solvation shell around the ion. Such contact ion pair solvation structures are found to be most stable in both MD and the DFT simulations on $DMP(H_2O)_N M^{x+}$ clusters for $Ca^{2+}$ and $Mg^{2+}$ ions (characterized by $\alpha \sim 152 - 173°$, cf. Table 2 and Fig. 5d,f).

## 3.4 Discussion

The presented linear infrared, femtosecond infrared pump-probe and absorptive 2D-IR spectra of the asymmetric phosphate stretching vibration $\nu_{AS}(PO_2)^-$ of DMP in aqueous solution in presence of 2M $Na^+$, $Ca^{2+}$ or $Mg^{2+}$ reveal clear differences between the different ions, supported by findings from MD simulations. The arising blue shifted signatures are compellingly assigned to the formation of contact ion pairs, on the order of 10-35 % of the total DMP concentration (16). Spectral differences among ions are enhanced in the nonlinear spectra yielding a higher resolution towards the detection of contact ion pair species while the linear spectra allow to quantify to some extent the presence of solvent separated ion pairs (15). While the 2D-IR spectra allow for a clear separation of different spectral components, the amplitudes of the femtosecond infrared pump-probe signals of the $Ca^{2+}$ or $Mg^{2+}$ samples appear distorted because of the spectral overlap of excited state absorption (positive pump-probe signal) the absorption decrease (negative pump-probe signal) on the fundamental transition of the oscillators of different species. Differences among DMP and DMP in contact with $Ca^{2+}$ or $Mg^{2+}$ are further revealed in the longer vibrational lifetimes $\nu_{AS}(PO_2)^-$ in contact ion complexes with $Ca^{2+}$ or $Mg^{2+}$. Possible reasons for the changes in vibrational lifetime are (i) the increased detuning to an acceptor (Fermi) resonance state (40) or (ii) the altered fluctuation properties of the environment experienced by the asymmetric $(PO_2)^-$ oscillator upon formation of contact ion pairs (cf. Fig. 5f). The largely preserved lineshape of the different spectral components, as revealed in the 2D-IR spectra of DMP with and without $Ca^{2+}$ and $Mg^{2+}$ ions points to (i) as the relevant mechanism for the observed increase of the vibrational lifetime. Due to the longer lifetime of $\nu_{AS}(PO_2)^-$ of the contact ion pair (cf. Fig. 2), a relative enhancement of the blue-shifted component is observed in the 2D signals.

The enhanced spectral separation supported by the lifetime differences of contact ion pair and DMP in neat $H_2O$ is particularly evident from the comparison of the cuts along the frequency diagonal (Figs. 3g,h). Nonlinear 2D-IR spectroscopy gives insight into the presence or absence of molecular couplings and accentuates differences of the different spectral features due to the different scaling of vibrational transition moments compared to compared to linear spectra ($\mu^4$ vs. $\mu^2$). The different aspects collectively are advantageous to the 2D-IR technique thus allowing for a much better separation and culminate in enhanced resolution to identify contact ion pair species of $Ca^{2+}$ and $Mg^{2+}$ with phosphate groups. Due to the subpicosecond lifetime of $\nu_{AS}(PO_2)^-$ the observation times in the 2D-IR measurements are essentially limited to the few-picosecond time scale. We have no indication for ion and/or water exchange on this time scale which would be reflected in a reshaping of the respective 2D lineshapes and the occurrence of cross peaks. These findings are consistent with the reported slow exchange of water molecules in the solvation shell around phos-phate groups (13, 41, 42) and the even longer residence time of ions (14). A quantitative analysis of the linear infrared absorption and 2D-IR spectra allows for estimating the fraction of DMP molecules which form contact ion pairs. As discussed in detail in Ref. 15 (supplement), 20-35% of all DMP molecules form contact pairs with $Mg^{2+}$ and a fraction of 12-30 % forms contact pairs with $Ca^{2+}$ for a total DMP concentration of 0.2 M and an ion concentration of 2 M.

The MD trajectories cover the microsecond time range for each investigated ion and start from contact pair geometries. We observe frequent, a moderate number and no transitions between contact ion pairs and solvent separated ion pairs for $Na^+$, $Ca^{2+}$ and $Mg^{2+}$, respectively. The characterization of geometric and energetic properties of the contact ion pairs relies on classical MD simulations employing fixed-charge force fields that only facilitate the presented microsecond simulation times. The theoretical description of ion pairing with the different ions $Na^+$, $Ca^{2+}$ and $Mg^{2+}$ is challenging in simulations due to uncertainties on quality of force fields, in particular for highly charged divalent cations, and further difficulties arise in obtaining converged results on long time scales, required, e.g, for the 2D representation of PMF. In particular, substantial recent efforts have been devoted to the improvement of ion parametrization of divalent cations $Ca^{2+}$ and $Mg^{2+}$ (37, 38, 43, 44, 45, 46). The employed 12-6-4 Li-Merz ion parameters mimic charge-induced dipole interaction and in the case of $Mg^{2+}$ provide superior results for a wide range of properties than commonly employed ion parameters (37). In particular, the combination of the 12-6-4 Li-Merz ion parameters and the TIP4P-FB water model was chosen due to a reduced tendency of contact ion pair formation of $Na^+$ and $Ca^{2+}$ with DMP.

To judge the reliability of the MD simulations we presented a comparison of geometric and energetic properties to DFT simulations of $DMP(H_2O)_N M^{x+}$ clusters for all three ions that inherently account for polarization and charge transfer effects. The comparison shows that geometric properties obtained from MD simulations are reasonably described, in particular the minimum $P\text{-}Ion^{x+}$ distance and the $P..O1..Ion^{x+}$ angle $\alpha$ show close agreement for the different contact ion pair geometries with the ions $Na^+$, $Ca^{2+}$ and $Mg^{2+}$. The findings are corroborated by Ref. (37) showing good correlation between gas-phase minimum and equilibrium contact distances extracted from free energy profiles. The comparison of the relative energetics of contact ion pair geometries in MD and DFT simulations on $DMP(H_2O)_N M^{x+}$ clusters further gives a consistent ordering of the energetics of observed contact ion species. We thus conclude that geometric and energetic properties derived from MD simulations employing 12-6-4 Li-Merz ion parameters in combination with the TIP4P-FB water model provide a reliable description of the contact ion pair geometries for the investigated ions $Na^+$, $Ca^{2+}$ and $Mg^{2+}$. Particular challenges are posed to the simulation of free energy differences of solvent separated and contact ion pairs for divalent cations and negatively charged biological moieties that can suffer from an overbinding of contact ion pairs. The respective 1D-PMF for $Mg^{2+}$ was given in Ref. (38) (12-6-4 Li-Merz ion parameters with TIP4Pew water model) and show a substantial stabilization of contact ion pairs compared to solvent separated ion pairs.

We note that the presented linear and 2D-IR measurements reveal that the simulated PMF (and underlying force field parameters) are not fully consistent. That currently precludes the quantitative description of stability of solvent separated and contact ion pairs for the divalent cation $Mg^{2+}$. Our experimental observations of contact ion pair formation by 20-35 % of DMP molecules with $Mg^{2+}$ and 12-30 % with $Ca^{2+}$ (15) introduce novel boundary conditions on the relative energetics of solvent separated and contact ion pairs. An extension of the approach to temperature-dependent measurements could provide access to ion-pairing enthalpies and entropies, to be compared with MD simulations. We consider the presented infrared data a benchmark and boundary condition for future ion model improvements, from which spectroscopic observables can be derived for a direct comparison between theory and experiment.

The peculiar differences of ion pairs formed by the $(PO_2)^-$ group and $Na^+$, $Ca^{2+}$, and $Mg^{2+}$ ions revealed in linear infrared, femtosecond infrared pump-probe and absorptive 2D-IR spectra were microscopically traced back to subtle differences in solvation structure around the ion and the $(PO_2)^-$ group. In particular, $Na^+$ ions replace a water molecule in the tetrahedral solvation environment around the $(PO_2)^-$ group but the solvation shell of the latter is largely preserved, thereby imposing moderate impact on the vibrational spectra of the $v_{AS}(PO_2)^-$ mode. $Mg^{2+}$ ions in contrast possess a tight and rigid octahedral solvation shell. In contact ion pairs with the $(PO_2)^-$ group, the latter can replace a single water in the ion solvation shell. Such geometrical arrangement directly correlates with the microscopic mechanism leading to the observed blue shift in the vibrational spectra of the $v_{AS}(PO_2)^-$ stretching vibration of DMP in aqueous solution (15,16). The partial de-solvation of the $(PO_2)^-$ group and the displacement asymmetric $(PO_2)^-$ mode vector brings the $(PO_2)^-$ group oxygen atom into close vicinity to the ion thereby accessing the repulsive part of interaction potential which leads to the observed substantial blue shifting of the $\nu_{AS}(PO_2)^-$ vibration. Contact ion pairs of the $(PO_2)^-$ group with $Ca^{2+}$ resemble the $Mg^{2+}$ situation but show substantially higher flexibility of solvation shell and larger ion radius.

## 4 Conclusions

We have presented a detailed analysis of linear infrared, femtosecond infrared pump-probe and absorptive 2D-IR spectra of the asymmetric phosphate stretching vibration of DMP, an established model system of the sugar-phosphate backbone of DNA and RNA, in presence of the ions $Na^+$, $Ca^{2+}$ and $Mg^{2+}$. The spectra reveal clear signatures of contact ion pair formation. Such effects underline the potential of nonlinear 2D-IR spectroscopy as an analytical probe of ion pair formation with phosphate groups. The experimental results are corroborated by a microscopic interpretation based on MD simulations, in turn benchmarked by density functional simulations on phosphate-ion-water clusters. The study reveals subtle structural differences of ion pairs formed by the $(PO_2)^-$ group and ions $Na^+$, $Ca^{2+}$ and $Mg^{2+}$ that are reflected in the highly sensitive vibrational response of the asymmetric phosphate stretching vibration. Currently, there is no consensus on the existence and amount of direct contact ion pairs, especially with biologically relevant $Mg^{2+}$ and $Ca^{2+}$ ions, and their impact on stability of macromolecular structures of DNA and RNA, because such systems are challenging to probe, both experimentally and through simulations. The present work addresses this topic of biological relevance with the help of a model system and brings new highly-needed data and insight into this debate.

**Acknowledgements:** This research has received funding from the European Research Council (ERC) under the European Union's Horizon 2020 research and innovation program (grant agreements No. 833365 and No. 802817). B. P. F. acknowledges support by the DFG within the Emmy-Noether Program (Grant No. FI 2034/1-1). We thank Janett Feickert for expert technical support.

**Table 1:** Time constants $\tau$ (in fs) of the $\nu=1$ population decay of the asymmetric phosphate stretching mode of DMP and DMP with 2 M excess ions. The mono-exponential numerical fits cover a delay range up to 3 ps.

|  | 1175 cm$^{-1}$ | 1220 cm$^{-1}$ | 1250 cm$^{-1}$ |
|---|---|---|---|
| DMP | 370 ± 50 | 340 ± 30 | 320 ± 30 |
| 2M Na$^+$ | 370 ± 50 | 340 ± 30 | 300 ± 30 |
| 2M Ca$^{2+}$ | 410 ± 50 | 350 ± 30 | 580 ± 30 |
| 2M Mg$^{2+}$ | 380 ± 50 | 300 ± 30 | 430 ± 30 |

**Table 2:** Comparison of contact ion pair geometric parameters for the different ions Na$^+$, Ca$^{2+}$ and Mg$^{2+}$. Geometric parameters from MD simulations (P....Ion$^{x+}$ distance and angle $\alpha = \sphericalangle(\text{Ion}^{x+}\ldots\text{O1}\ldots\text{P})$) are derived for the respective contact ion pair (CIP) minimum of the 2D-PMF depicted in Figure 5, O1...Ion$^{x+}$ distance is derived from the first maximum of the respective rdf with the full-width-half-maximum given in parenthesis. Geometric parameters of the DMP(H$_2$O)$_n$M$^{x+}$ clusters are derived from respective minima with vibrational frequencies $\nu_{AS}(\text{PO}_2)^-$ = 1208, 1226 and 1248 cm$^{-1}$ for DMP(H$_2$O)$_{11}$Na$^+$ , DMP(H$_2$O)$_{13}$Ca$^{2+}$, and DMP(H$_2$O)$_{11}$Mg$^{2+}$, respectively, DFT optimized cluster structures taken from Ref. (16). $\Delta E$ denotes the free energy difference taken along the angular coordinate of the 2D-PMF and the energetic difference of minimal energy cluster structures obtained via constrained optimization along the angular coordinate $\alpha = \sphericalangle(\text{M}^{x+}\ldots\text{O1}\ldots\text{P})$ which allows to interpolate between the tetrahedral hydrogen bond geometry around the O1 atom of the (PO$_2$)$^-$ and the prototypical solvation geometry with octahedral coordination of the ion in the minimal cluster models DMP(H$_2$O)$_n$M$^{x+}$. Positive $\Delta E$ favor the tetrahedral hydrogen bond geometry while negative $\Delta E$ favor the intercalation of DMP-oxygen atoms into the solvation shell around the ion via replacement of a water molecule (cf. Fig. 5 e-f).

|  | MD | DMP(H$_2$O)$_n$M$^{x+}$ |
|---|---|---|
| **Na$^+$:** |  | **DMP(H$_2$O)$_{11}$Na$^+$** |
| P....Na$^+$ | 3.44 Å | 3.58 Å |
| O1...Na$^+$ | 2.23 Å [2.15 – 2.37] | 2.45 Å |
| $\alpha = \sphericalangle(\text{Na}^+\ldots\text{O1}\ldots\text{P})$ | 132.8° | 127.3° |
| $\Delta E$ | 396.3 cm$^{-1}$ | 1069 cm$^{-1}$ |
|  |  |  |
| **Ca$^{2+}$:** |  | **DMP(H$_2$O)$_{13}$Ca$^{2+}$** |
| P....Ca$^{2+}$ | 3.78 Å | 3.66 Å |
| O1... Ca$^{2+}$ | 2.33 Å [2.27 – 2.42] | 2.25 Å |
| $\alpha = \sphericalangle(\text{Ca}^{2+}\ldots\text{O1}\ldots\text{P})$ | 163.2° | 152.2° |
| $\Delta E$ | -285.8 cm$^{-1}$ | -218.2 cm$^{-1}$ |
|  |  |  |
| **Mg$^{2+}$:** |  | **DMP(H$_2$O)$_{11}$Mg$^{2+}$** |
| P....Mg$^{2+}$ | 3.48 Å | 3.47 Å |
| O1... Mg$^{2+}$ | 2.03 Å [1.97 – 2.08] | 1.97 Å |
| $\alpha = \sphericalangle(\text{Mg}^{2+}\ldots\text{O1}\ldots\text{P})$ | 167.2° | 173.9° |
| $\Delta E$ | -2645 cm$^{-1}$ | -1086 cm$^{-1}$ |

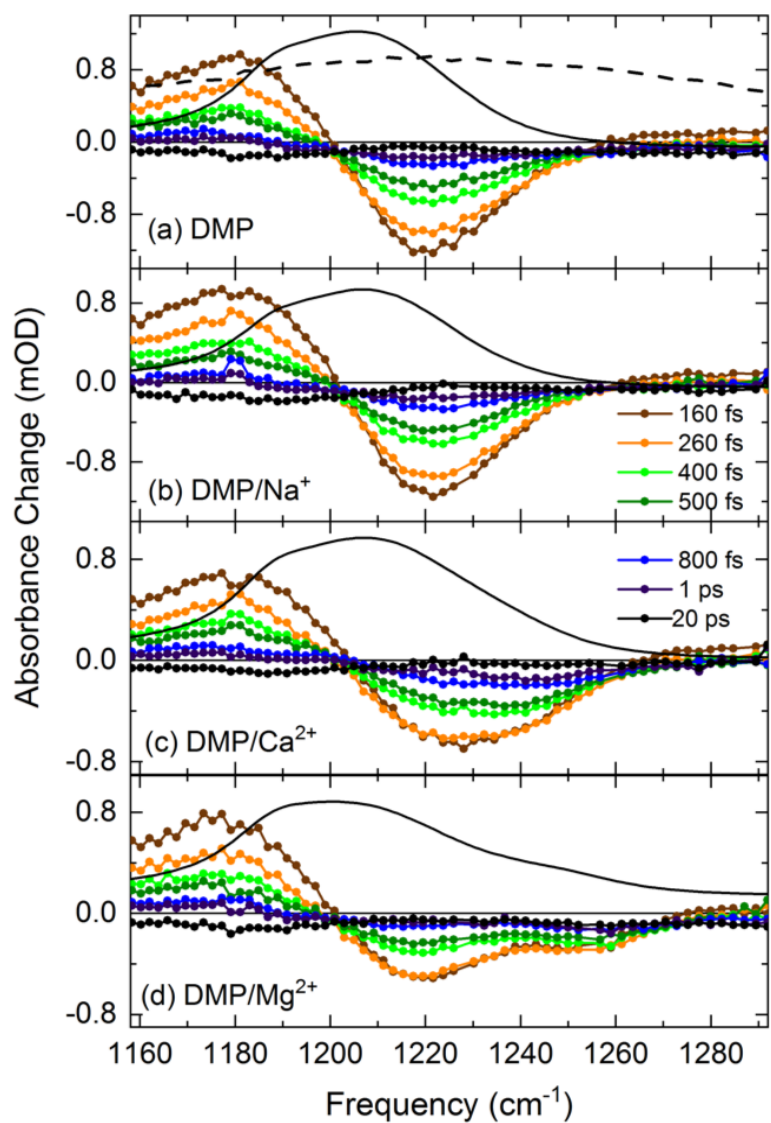

**Fig. 1:** Femtosecond pump-probe spectra in the range of the asymmetric phosphate stretching band for (a) DMP in $H_2O$, (b) DMP in $H_2O$ with 2 M $Na^+$ (c) DMP in $H_2O$ with 2 M $Ca^{2+}$, and (d) DMP in $H_2O$ with 2 M $Mg^{2+}$ after excitation with a pump pulse centered at 1220 $cm^{-1}$ (dashed line in panel (a): pump spectrum). The change of absorbance $\Delta A = -\log(I/I_0)$ is plotted as a function of probe frequency for the given delay times ($I$, $I_0$: probe intensity transmitted through the sample with and without excitation). The black solid lines represent the respective linear infrared absorption spectrum.

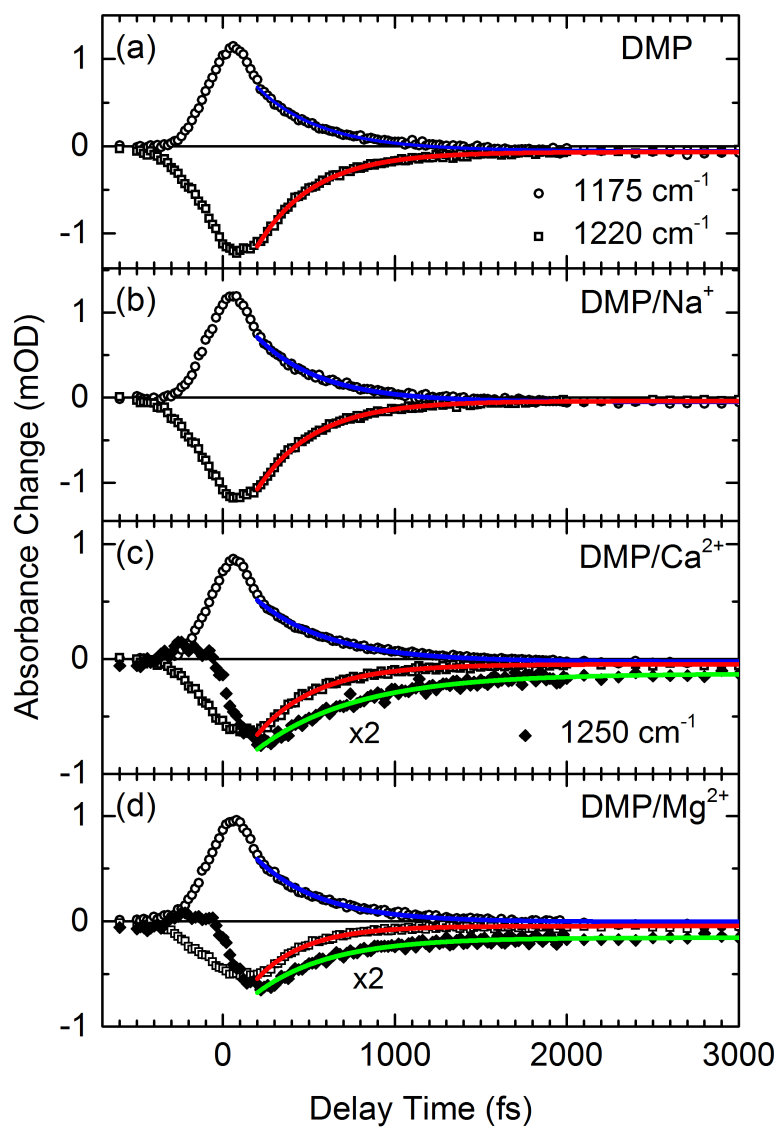

**Fig. 2:** Time resolved pump-probe signals at fixed probe frequencies of 1175 cm$^{-1}$ (open circles), 1220 cm$^{-1}$ (open squares), and 1250 cm$^{-1}$ (diamonds) as a function of delay time for the four samples. The solid colored lines are single exponential numerical fits to the transients. The derived decay times are summarized in Table 1. The amplitudes of the kinetic traces at 1250 cm$^{-1}$ (panels (c) and (d)) have been multiplied by a factor of 2.

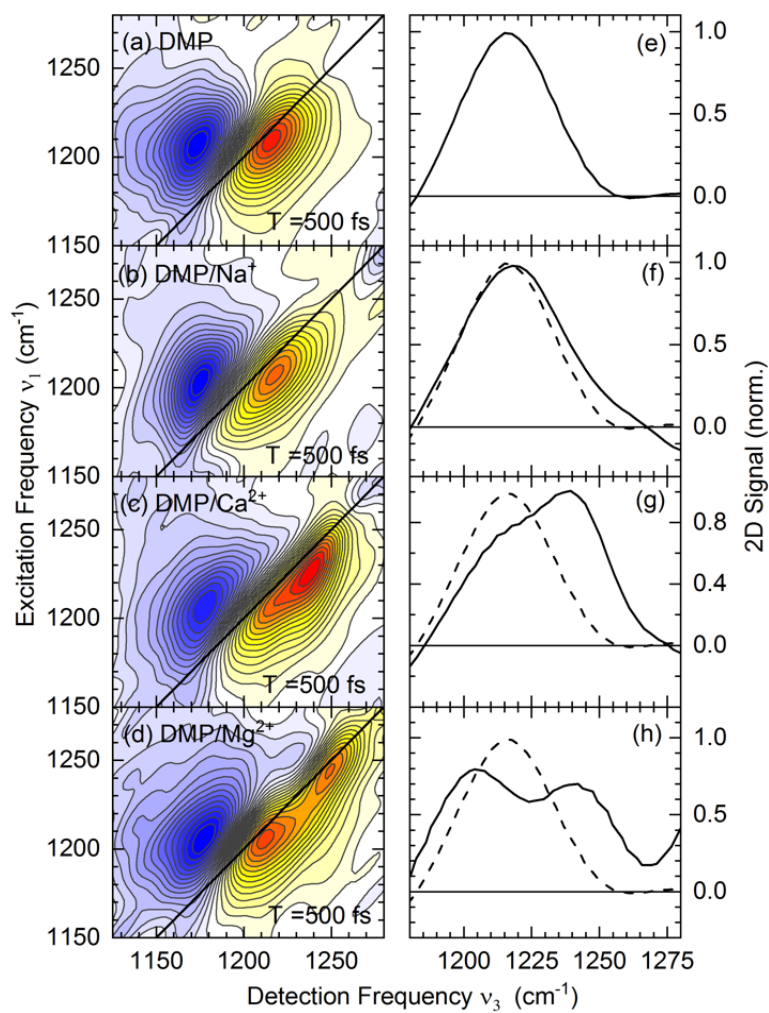

**Fig. 3:** 2D-IR spectra of the asymmetric phosphate vibration measured at a waiting time T=500 fs. (a) DMP in $H_2O$, (b) DMP in $H_2O$ with 2 M $Na^+$, (c) DMP in $H2O$ with 2 M $Ca^{2+}$, and (d) DMP in $H_2O$ with 2 M $Mg^{2+}$. The yellow-red contours show the absorption decrease on the v=0-1 transition, the blue contours the absorption increase on the v=1-2 transition. (e-h) Frequency cuts of 2D-IR spectra along a diagonal crossing the maximum of the respective v=0-1 peak (solid lines). The dashed line in (f-h) represents the diagonal cut of the 2D-IR spectrum of DMP in water for reference.

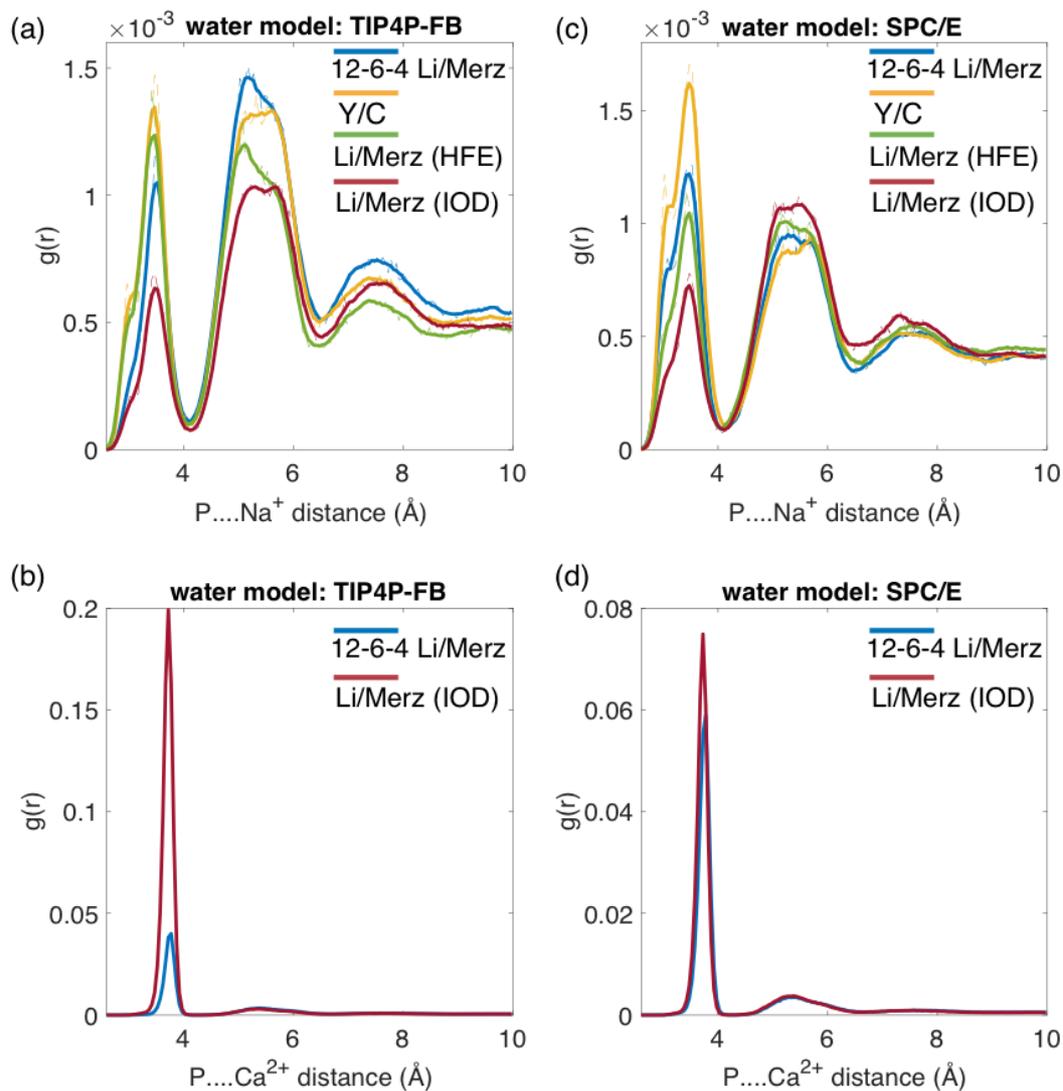

**Fig. 4:** (a,b) Radial distribution functions (rdf) of the DMP P…Na$^+$ distance for different combinations of water model and ion force field. Simulations times are 0.99 μs for 12-6-4 Li/Merz, 0.98 μs for Young/Cheatham (Y/C), 0.99 μs for Li/Merz (HFE and IOD) ion parameters employing the TIP4P-FB water model and 0.79 μs for 12-6-4 Li/Merz, 0.79 μs for Young/Cheatham, 0.79 μs for Li/Merz (HFE) and 0.99 μs for Li/Merz (IOD) ion parameters employing the SPC/E water model. Solid lines give the ±0.1 Å moving average of g(r) (dashed line). (c,d) Radial distribution functions (rdf) of the DMP P…Ca$^{2+}$ distance for different combinations of water model and ion force field. Simulations times are 1.59 μs for 12-6-4 Li/Merz and 0.79 μs for Li/Merz (IOD) ion parameters employing the TIP4P-FB water model and 0.79 μs for 12-6-4 Li/Merz and Li/Merz (IOD) ion parameters employing the SPC/E water model.

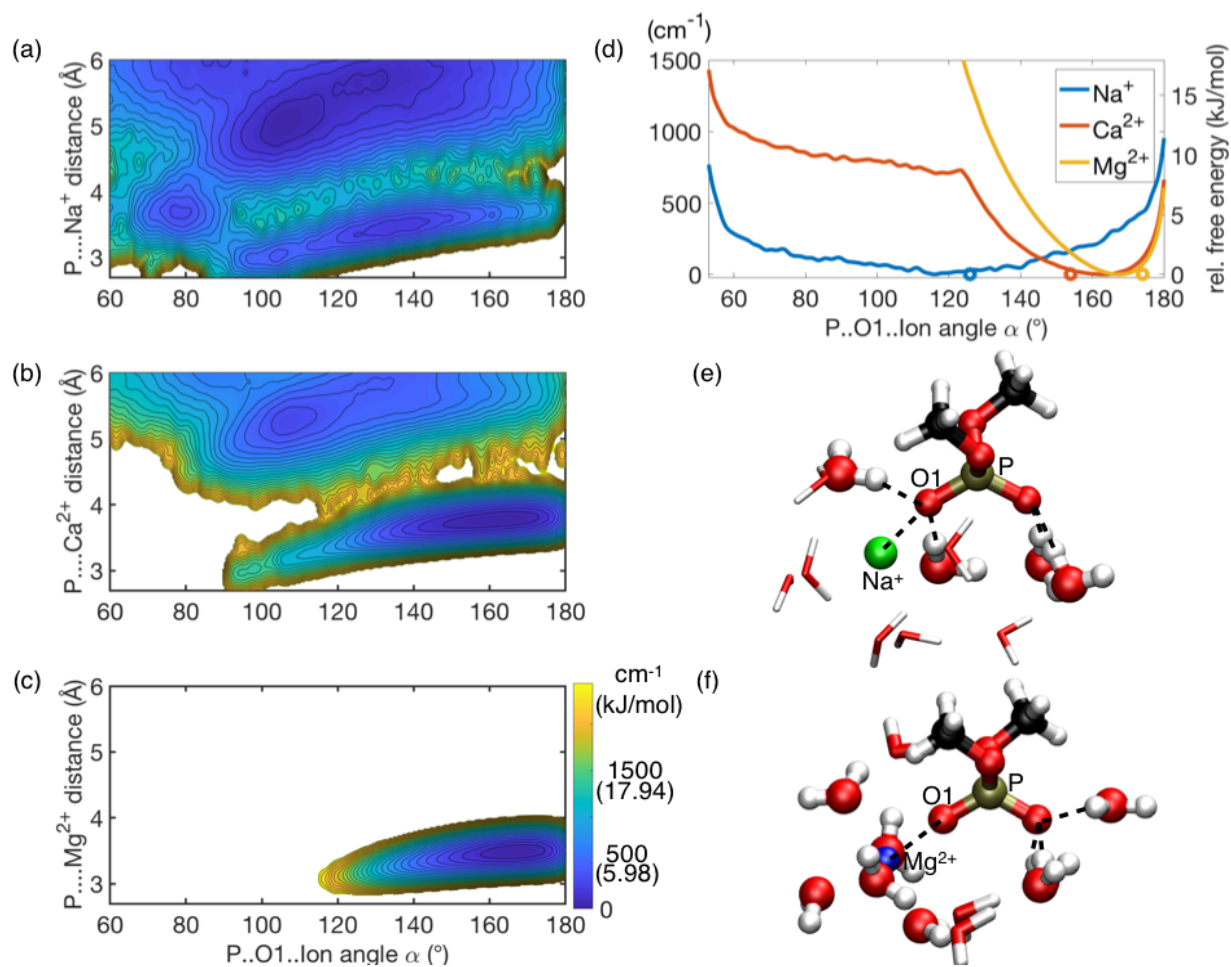

**Fig. 5:** Two dimensional potential of mean force (2D-PMF) along the P…ion distance and the angular coordinate α = ∢(M$^{x+}$…O1…P) obtained from 1.59 μs molecular dynamics trajectories (M$^{x+}$ = Na$^+$, Ca$^{2+}$, Mg$^{2+}$). White areas in the plots are not sampled by the respective system. (d) Angular dependence of the free energy. Minimum energy angular energy profiles were derived from the 2D-PMF upon locating the energy minima for each α for contact ion pair P…ion distances (< 4 Å). Colored symbols mark the α-values of the DMP(H$_2$O)$_N$M$^{x+}$ cluster minimum geometries (with M$^{x+}$ = Na$^+$, Mg$^{2+}$: N = 11 and M$^{x+}$ = Ca$^{2+}$: N = 13). (e) Prototypical solvation geometry of the DMP(H$_2$O)$_{11}$Na$^+$ cluster with α~127° and the Na$^+$ ion shown in green. The Na$^+$ ion intercalates into the tetrahedral hydrogen bond geometry around the O1 atom of the (PO$_2$)$^-$ group. Hydrogen bonded water molecules are shown in ball-and-stick representation, water molecules of the Na$^+$ solvation shell in stick representation. (f) Prototypical solvation geometry of the DMP(H$_2$O)$_{11}$Mg$^{2+}$ cluster for angle α~174° with the Mg$^{2+}$ ion shown in blue. Water molecules in the octahedral solvation shell around Mg$^{2+}$ and hydrogen bonded water molecules are in ball-and-stick representation, bridging water molecules in stick representation. The O1 atom of the (PO$_2$)$^-$ group takes the position of one of the water oxygen atoms in the octahedral solvation geometry around the Mg$^{2+}$ ion. DFT optimized cluster structures are taken from Ref. (16).